\documentclass[10pt, conference, letterpaper]{IEEEtran}
\usepackage{endnotes,epsfig}

\usepackage{amssymb}
\usepackage{amsmath}
\usepackage{amsfonts}
\usepackage{url,color,fancyvrb}
\usepackage{times}
\usepackage{subfigure}
\usepackage{array}
\usepackage{xspace}
\usepackage{multirow}
\usepackage{multicol}
\usepackage[ruled]{algorithm2e}
\usepackage{algpseudocode}
\usepackage{enumitem}
\usepackage{listings}
\graphicspath{{./figures/}}

\newcommand{\para}[1]{\vspace*{1ex}\noindent\textbf{#1}}

\newcommand{\authnote}[2]{\ifnum\authnotes=1\begin{quote}\textbf{#1 says:} #2\end{quote}\fi}

\newcommand{\void}[1]{{}}

\newcommand{\mysys}{{IOArbiter}\xspace}
\newcommand{\stimpl}{\emph{storage implementation}\xspace}
\newcommand{\stdb}{{\emph{StateDatabase}}\xspace}
\newcommand{\stsched}{{\emph{Scheduler}}\xspace}
\newcommand{\stbroker}{{\emph{StorageBroker}}\xspace}
\newcommand{\stmgr}{{\emph{StorageManager}}\xspace}
\usepackage[utf8]{inputenc}
\usepackage{cleveref}
\usepackage{gensymb}

\newcommand{\SKELETON}[1]{{\em #1}}
\renewcommand{\SKELETON}[1]{}		

\newcommand{\mytitle}{\mysys: Dynamic Provisioning of Backend Block Storage in the Cloud} 

\begin{document} 

\title{\mytitle}

\author{
Moo-Ryong Ra and Hee Won Lee\\
AT\&T Labs - Research}

\maketitle 

%
\begin{abstract}
With the advent of virtualization technology, cloud computing realizes on-demand computing.
The capability of dynamic resource provisioning is a fundamental driving factor for users 
to adopt the cloud technology. The aspect is important for cloud service providers 
to optimize the expense for running the infrastructure as well.
Despite many technological advances in related areas, however, 
it is still the case that the infrastructure providers must decide hardware configuration 
before deploying a cloud infrastructure, especially from the storage's perspective. 
This static nature of the storage provisioning practice can cause many significant problems 
in meeting tenant requirements, which often come later into the picture.
In this paper, we propose a system called \mysys that enables the dynamic creation of 
underlying storage implementation in the cloud. 
\mysys defers storage provisioning to the time at which a tenant actually requests a storage space.
As a result, an underlying storage implementation, e.g., RAID-5/6 or Ceph storage pool 
with (6,3) erasure coding, will be materialized at the volume creation time. 
%
Using our prototype implementation with OpenStack Cinder, we show that \mysys can simultaneously 
satisfy a number of different tenant demands, which may not be possible with a static configuration.
Additionally the built-in QoS mechanisms, including admission control and dynamic throttling,  
help \mysys system mitigate a noisy neighbor problem among tenants.

\end{abstract}
\section{Introduction}
\label{sec:intro}

With the advent of virtualization technology, cloud computing materializes 
on-demand computing. A tenant can create a virtual infrastructure based 
on their exact needs in a much shorter amount of time and the amount of 
provisioned resources can be adjusted anytime from anywhere by the tenant. 
The capability of dynamic resource provisioning is a fundamental driving 
factor for users to adopt the cloud technology. The aspect is also important 
for cloud infrastructure providers in order to optimize the overall expense 
(CapEx/OpEx) for running the infrastructure.

For enterprise storage solutions, vendors drove an appliance-based model that
typically provide proprietary software and customized hardware 
to support a specific class of applications, e.g., database, and/or 
a specific layer of storage stack, e.g., NFS storage, SAN block storage, etc.
However, since many different tenant/storage applications will share the same 
infrastructure in the cloud setting, the underlying storage implementation should 
be able to be adapted for active tenants.
Unfortunately, despite the fact that many recent technological advances 
have been made in virtualizing compute, network, and storage resources in the cloud, 
it is still the case that the infrastructure providers should decide 
hardware configuration before deploying a cloud infrastructure, especially from the storage's perspective. 
For instance, the OpenStack Cinder~\cite{OpenStackCinder} service, 
which governs block storage management tasks, mandates a preconfigured 
\stimpl\footnote{In this paper, we use the term
\emph{``storage implementation''} as a software/hardware configuration 
on top of which a logical volume will be created. 
For instance, it can be a traditional RAID if one uses local disks
for cloud block storage, or a Ceph storage pool if one relies upon such a system.}
before the operation, 
e.g., through local RAID configuration, vendor appliances/solutions, etc. 
In a multi-tenant cloud environment, in particular, the static nature of 
the storage provisioning practice can cause many significant problems 
for meeting tenant application requirements, which often come later into the picture.

\begin{table}[t]
\centering
\small
\begin{tabular}{|m{4.0cm}|m{2.2cm}|m{1.8cm}|} \hline
\textbf{Application} & \textbf{A single \stimpl} & \textbf{Dynamic provisioning} \\ \hline
\hline
VDI	that requires 3x redundancy from underlying storage systems	& 3x 	& 3x 	\\ \hline
vCDN that implements 3x redundancy in the application layer.	& 9x 	& 3x 	\\ \hline
\hline
Total Storage Overhead & 12x & 6x			\\ \hline
\end{tabular}
\vspace{0.5ex}
\caption{
\label{tbl:introex}A case of an inefficient utilization of storage capacity.}
\vspace{-4ex}
\end{table}

Table~\ref{tbl:introex} shows an illustrative example. We consider 
two applications. One is a Virtual Desktop Infrastructure (VDI) application and provides a 
virtual desktop environment for corporate employees through VMs in the cloud.
It requires reliable storage, and consequently demands triple (3x) replication strategy, 
which is a common industry practice.
The other application, vCDN, is a virtualized Content Delivery Network (CDN)
application. Unlike the VDI application, vCDN implements triple redundancy in its application layer. 
In this case, since the application replicates data across three data centers, 
it is not necessary for underlying storage systems to provide any additional redundancy.
Suppose that we first deployed the VDI application to the cloud, and after some time 
deploy the vCDN application in the same cloud infrastructure. 
Since the architects only knew about the VDI application when they design the infrastructure,
they implemented storage systems based on the triple replication strategy.
Even though the vCDN application, which comes later into the picture, does not 
require an additional redundancy, it needs to be deployed into the same infrastructure. 
As a result, we end up with wasting a lot of storage space, i.e., 12x vs. 6x in the table.
In many practical scenarios, it is not easy to avoid this kind of situation due to the 
rigid nature of storage services.
However, if we can somehow configure the underlying storage parameters dynamically, e.g., 
this problem might be resolved as depicted in Table~\ref{tbl:introex}.
In this particular example, total storage overhead of a single \stimpl is 
twice as much as the ideal case.

In this paper, we propose a system called \mysys that enables the 
dynamic creation of underlying \stimpl in today's cloud. 
\mysys defers the implementation of underlying storage to the volume creation time, 
i.e., the time at which a tenant actually requests a storage space.
To avoid humongous design space, a cloud infrastructure provider 
may define a customized set of storage implementation types 
so as to incorporate a range of performance and reliability levels, 
e.g., RAID-5 or 6 with a minimum of 200 IOPS, 
Ceph with (6, 3) or (10, 4) erasure coding, etc. and use them with \mysys system.
%
When a tenant request, e.g., volume creation, is come into the system,
\mysys analyzes the request and automatically creates a necessary 
\stimpl if it is not yet available in the infrastructure, 
e.g., RAID-5 with 6 disks or Ceph with (10, 4) erasure coding, etc. 
%
As a management layer of cloud block storage services, \mysys has
a number of useful features, including
a) an ability to perform garbage collection, e.g, reclaiming unused space 
and/or a \stimpl, 
b) an admission control and dynamic throttling mechanism that enables 
per-VM IOPS allocation,
and
c) a containerized control plane for effective maintenance.

%
We implemented an \mysys prototype with OpenStack Cinder~\cite{OpenStackCinder}.
Our preliminary evaluation result shows that 
\mysys can simultaneously satisfy heterogeneous tenant demands, 
which may not be even possible with a static \stimpl. 
Moreover, \mysys has a negligible overhead in terms of scheduling operation.

The rest of the paper is organized as follows. Section~\ref{sec:motiv} gives 
some basic information on OpenStack Cinder and further motivates the problem. 
Section~\ref{sec:sysarch} then describes design rationales and overall system architecture.
The detailed description of how we implement our system with OpenStack is presented 
in Section~\ref{sec:impl}. We provide our preliminary evaluation results with our prototype system
in Section~\ref{sec:eval}, our investigation on related work in Section~\ref{sec:related}, 
and conclude the paper.


%
\section{Background and Motivation}
\label{sec:motiv}

In this section, we provide a brief overview on 
the block storage management layer in the cloud
and present several exemplary problems caused by 
the static nature of storage provisioning practices.  

\subsection{Cinder: virtualized block storage management layer}
In this paper, we assume a cloud environment based on OpenStack 
open source software in order to manage underlying physical computing resources.
There exist many alternatives including 
both proprietary~\cite{aws, azure, googlecloud} and open source solutions~\cite{cloudstack}. 
Nonetheless, OpenStack is chosen by us since it is not only an open source solution, 
but is also supported by a large community behind it.
OpenStack has a block storage management layer, 
Cinder~\cite{OpenStackCinder}, 
which is a primary focus of this paper. 
Cinder governs a control path of cloud block storage, e.g., creating/deleting 
logical volumes for tenant VMs on the provisioned \stimpl, 
connecting/disconnecting them to/from VMs, etc.

Cinder service is comprised of several sub-components. 
The \emph{cinder-scheduler} typically runs on the OpenStack cluster's controller nodes.
When a block storage request is issued by users through REST API calls, 
\emph{cinder-scheduler} determines which storage node should handle a given request (e.g., volume creation/deletion). 
In each storage node, one or more \emph{cinder-volume} services will be instantiated, take the 
request object, and actually perform the designated task. 
All communications among \emph{cinder-scheduler} and \emph{cinder-volume} go through
a common message bus, e.g., RabbitMQ~\cite{rabbitmq}, ZeroMQ~\cite{zeromq}, etc.
When \emph{cinder-volume} receives a request, manipulating logical volumes 
such as creating/deleting a volume is performed through a \emph{volume-driver} 
specific to the underlying \stimpl.
For instance, a RAID configuration based on local disks
typically relies upon the LVM driver implementation. 
Some open source distributed storage systems provide their own drivers, e.g., Ceph RBD. 
Vendor appliances have their own driver to properly interact 
with proprietary software and hardware.
For the plumbing part, i.e., connecting/disconnecting a logical volume to/from VMs,
Cinder also supports different types of transports, e.g., iSCSI, Fibre channel, etc.
Additionally, Cinder provides a plug-in framework for \emph{cinder-scheduler} 
so that a developer can implement a customized \emph{scheduler filter}.
\mysys implements its own volume placement strategy 
using the \emph{scheduler filter} interface.

\subsection{Growing pains in cloud block storage: example cases}

In addition to the motivating example given in Section~\ref{sec:intro},
our organization have faced several problems due to the static nature 
of the storage provisioning practice. 
Here we illustrate some of them to further motivate the problem\footnote{Some of these stories 
are slightly modified from the real events for more accessible description.}.

\para{Reconfiguring storage.} Our IT team provides a disk array preconfigured 
with RAID-6 with arbitrary internal partitions.
There were no problems in the beginning. Later, however, one of our development teams
wanted to deploy a virtualized CDN application that replicates its data across multiple datacenters
in the application layer.
In this case, using RAID-6 for the application is a waste of storage space since it will be 
unnecessarily replicating data internally.
It was hard for the IT team to change it easily since other tenants are already using
the array.

\para{Meeting multiple performance requirements.} A storage array that has 24 SATA HDDs 
is configured as JBOD\footnote{Just a Bunch Of Disks} and used by one application that does not require 
redundancy on the storage layer. Later, another application comes to the cloud 
and requests a minimum of 500 IOPS\footnote{I/O operations per second} for its VMs' block storage traffic. 
Unfortunately, its workload is mainly composed of 4 KBytes random read/write, 50\% each.   
Since the entire disk array is configured as JBOD and a single disk can support $\sim$200 IOPS
for this type of workload. If the underlying storage implementation was RAID-5 or 6 with
multiple disks, the requirement might be supported more easily. So they have no other choice than to
construct a software RAID inside a VM. Unfortunately, however, the disks in the array were all 
partially occupied by other VMs. 

\para{Isolation from other tenants.} A storage array is configured as RAID-6 and 
multiple tenants start to utilize the space. After some time, a tenant from a 
government organization requests a physically isolated storage for their data. 
Despite the fact that there are sufficient storage space remained in the disk array, 
it was not trivial to carve out some of the disks for a new tenant given that SLAs 
require 24/7 uptime. 

The cases including all of the above-mentioned scenarios, but not limited to, can be benefited 
by \mysys-like systems that can dynamically configure and change 
the underlying \stimpl at runtime.

\section{\mysys Design}
\label{sec:sysarch}

\subsection{Design principle}

\mysys is designed based on two fundamental principles:
\emph{late binding} and \emph{non-intrusiveness}. 

\para{Late binding.} To enable dynamic provisioning of 
a \stimpl, \mysys adopts the \emph{late binding}
principle, which is a popular computer programming mechanism.
Inspired by this principle, a \stimpl will be bound to the cloud infrastructure
at the request time, i.e., volume creation time.
Storage medium may be provided as a bunch of raw disks
or with a state ready to be configured, e.g., 
a vanilla Ceph installation without any configured storage pools.
Then, any relevant \stimpl will be created when necessary, 
e.g., a RAID for the former and a storage pool for the latter example.

\para{Non-intrusiveness.} Since \mysys is aimed to be 
deployed in production clusters, it is desirable to minimize 
probable impacts on existing OpenStack components. 
We make a couple of important decisions for this purpose.
First, we implement \mysys as a filter and driver for the Cinder service
rather than making the system as a separate stand-alone service. 
With this approach, \mysys can naturally integrate with existing components,
and consequently minimize potential incompatibility problems.
Second, we exploit a container technology to isolate a newly created \stimpl. 
This decision ensures a dedicated agent per \stimpl, and 
thereby isolates the instance from other {\stimpl}s. As a result,
the system administrator can easily enable/disable the \stimpl~as needed.
Most failures in a given \stimpl will not affect other {\stimpl}s.

\subsection{System architecture}

\begin{figure}[t]
  \centering
  \includegraphics[viewport=55 130 750 570, scale=0.33, clip=true]{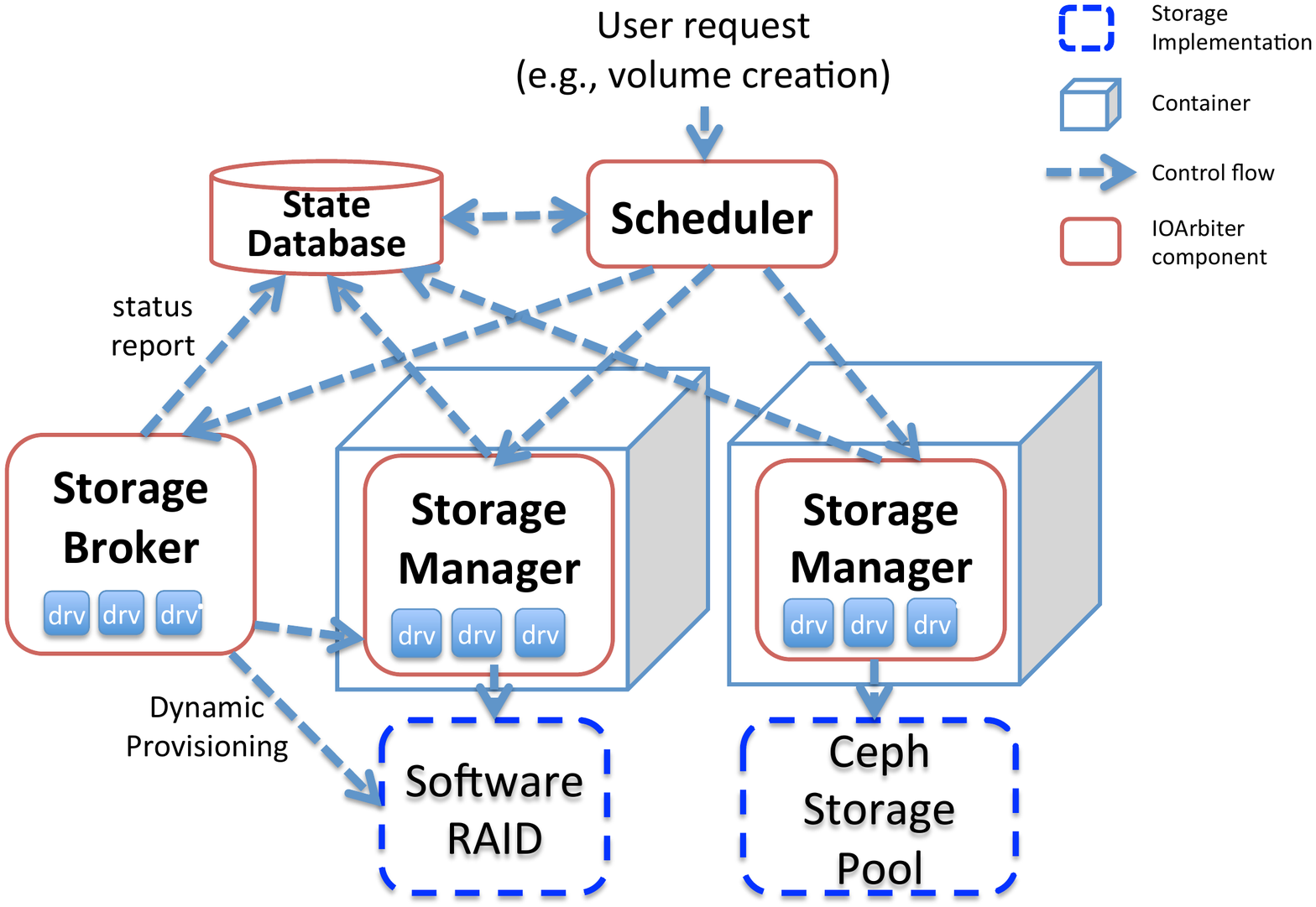}
  \caption {\mysys system architecture. Two probable {\stimpl}s are presented.}
  \label{fig:ioarbarch}
\end{figure}


Figure~\ref{fig:ioarbarch} illustrates \mysys system architecture.
The system is comprised of several components including \stsched, \stdb,
\stbroker and \stmgr.
When a user request arrives at the system with a storage specific requirement, 
e.g., a triple replicated volume with 100 minimum IOPS, 
\stsched first checks the resource availability
based on the data from \stdb and decides who should handle the request, 
i.e., either \stbroker or \stmgr. 
If a new \stimpl is required by the request due to the absence of resources, 
\stbroker will take the request and create a new \stimpl. 
Otherwise, i.e., if there exists a \stimpl that can support the request, 
\stsched will route the request to one of the available \stmgr.
A \stimpl can be a RAID configuration based on local disks, 
a storage pool based on a distributed storage systems like Ceph~\cite{Weil2006Ceph}, 
or any storage vendor solution. 
When \stbroker creates a \stimpl, it will additionally create a \stmgr associated 
with the \stimpl. \stmgr will execute actual storage commands, 
such as creating/deleting a logical volume at a \stimpl, 
attaching/detaching it from VMs, etc. 
It is worth noting that \mysys uses container technology to isolate the newly 
created control path of \stmgr.

Once instantiated, \stbroker and \stmgr instances will report their states 
to \stdb so that \stsched can perform the scheduling actions. 
The report message from \stbroker has a set of information about raw resources 
for which the \stbroker is responsible. For instance, it can be the number of
disks, disk types, and mediums in case of a software RAID-based \stimpl.
In contrast, the report from \stmgr contains more specific information of the \stimpl.
It includes the number of logical volumes already created in the \stimpl, total
performance budget and allocated resources, capacity status, etc.

\subsection{Performance isolation}
\mysys is intended to be deployed for multi-tenant cloud environments.
In this context, performance isolation is an important issue. 
\mysys provides two features -- admission control based on
offline profiles and dynamic throttling based on runtime characteristics. 

\para{Admission control.} When a new \stimpl is created, a containerized 
\stmgr will be created along with it. At that time, \mysys ensures that 
a new \stmgr has a total performance budget for that particular \stimpl.
For instance, if the \stimpl is based on a single SATA HDD,  
\mysys allocates $\sim$200 for the worst-case 4k-block IOPS budget.
After that, whenever a new volume creation request comes into the system,
\stmgr manages budgets and inform the remaining budgets to \stdb.
This will ensure that, if a budget is full for a given \stimpl, no more requests
will be issued to the corresponding \stmgr. 

\LinesNumbered
\SetInd{1ex}{2ex}
\begin{algorithm}[ht]
	\scriptsize
	Set \emph{minimum IOPS} for each volume $i$ ($R_{i}$)\;
	By default, $R_{i} = 0$\;
	Set control interval: M\;
	\BlankLine

 	\While {wait M seconds} {
		\emph{is\_throttling = false}\;
		\BlankLine
		
		\ForEach{volume} {
			collect \emph{current IOPS}\;
			\uIf {current IOPS $<$ minimum IOPS} {
				throttle each volume $i$ by $R_{i}$\;
				\emph{is\_throttling = true};
			}
			\Else {
				continue\tcc*[r]{do nothing}
			}
		}
		\BlankLine
		
		\uIf {is\_throttling == false} {
			release all throttling\;
		}
		\Else {
			continue\tcc*[r]{do nothing}
		}		
	}
	\caption{\small{Dynamic Throttling}}
	\label{algo:dynamic_throttling}
\end{algorithm}

\para{Dynamic throttling.} Although \mysys allocates performance, e.g., IOPS, based
on the budget, it might be the case that some resource interference still happens 
among the allocated volume traffic. \mysys provides a dynamic throttling mechanism
to mitigate the problem, as shown in Algo.~\ref{algo:dynamic_throttling}.
\mysys monitors a runtime statistics (i.e., \emph{current IOPS}) of each volume (at line 7) and, 
if the performance requirement (i.e., \emph{minimum IOPS})  of a certain volume is violated (at line 8), 
it suppresses other flows sharing the same \stimpl 
based on their worst-case requirements (at line 9). 
If no performance requirements are violated, 
throttling actions will be disabled (at line 16).

\subsection{Discussion}
%

\para{Garbage collection.}
Since \mysys encourages dynamic provisioning of \stimpl,
there exists a concern on resource fragmentation. 
To mitigate this problem, \mysys has a notion of periodic 
garbage collection mechanism.
Based on the mechanism, \mysys is able to reclaim unused storage space, if any, 
and/or rebalance the skewed data in already deployed distributed storage systems.

\para{High available and scalable service.}
%
\mysys system belongs to a control plane of cloud block storage.
In a large scale cloud infrastructure where 1000s of nodes or more can be deployed, 
it is important to make a service both highly available and scalable. 
Making a centralized gateway to be highly available, balancing incoming traffic load
uniformly across the available resources, and handling partitioned service resources
are classic topics of distributed systems. 
We do not attempt to make any contribution in the above-mentioned problem space.
Instead, \mysys could exploit an external service that is dedicated to these functions, 
such as Pacemaker~\cite{pacemaker}.

\section{Implementation of \mysys system}
\label{sec:impl}

\begin{figure}[t]
  \centering
  \includegraphics[viewport=55 180 750 570, scale=0.33, clip=true]{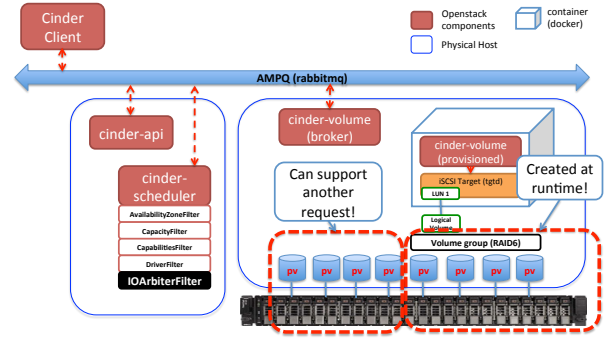}
  \caption {\mysys implementation with OpenStack Cinder}
  \label{fig:ioarbimpl}
\end{figure}

We have implemented an \mysys prototype with OpenStack Cinder service~\cite{OpenStackCinder}.
The integrated system is described in Fig.~\ref{fig:ioarbimpl}.
The \mysys sub-components are mapped onto Cinder components. 
A tenant request is described as a Cinder \emph{volume-type}, which can have 
a set of user-defined key-value pairs, e.g., \{redundancy=5,min-iops=100,iosize=4k,..\}.
Cinder has an RESTful API interface (\emph{cinder-api}) and the communication between
clients and \emph{cinder-api} is through a message queue service, e.g., 
RabbitMQ~\cite{rabbitmq}, ZeroMQ~\cite{zeromq}, etc.
\stsched is mapped onto \emph{cinder-scheduler} and implemented as one of its filters. 
\stbroker and \stmgr each are mapped onto \emph{cinder-volume} with a different mode 
of operations -- \emph{cinder-volume(broker)} and \emph{cinder-volume(provisioned)} --
and implemented as a separate driver\footnote{For instance, an extended version of LVM driver.}.
In Fig.~\ref{fig:ioarbimpl}, as an example, we present a software-based 
RAID configuration based on local disks available in a single storage node.
It should be noted that \mysys can support other types of \stimpl such as configuring 
a storage pool for Ceph, etc.

As discussed in Section~\ref{sec:sysarch}, \mysys exploits a container technology 
to isolate a newly created control path. We use the Docker container~\cite{docker} 
for our prototype implementation.
The \emph{cinder-volume} uses a local configuration file (cinder.conf) 
to configure and maintain its internal states. 
\mysys inserts an additional information in that file for our admission control mechanism.
When a containerized \emph{cinder-volume} (i.e., \emph{cinder-volume(provisioned)}) is created, \stbroker (i.e., \emph{cinder-volume(broker)}) 
will insert pre-computed performance budget into a cinder.conf file 
inside the container. Those budget numbers are based on offline profiles and specific to a given
\stimpl. After that, buget accounting operations are done by \emph{cinder-scheduler} based
on the reported information from the containerized \emph{cinder-volume} service.
Dynamic throttling service will be instantiated inside each container, run as a daemon process,
and periodically collect block device performance in case of software-RAID based \stimpl. 
When an SLA violation is observed, the service throttles all other storage traffic sharing
the \stimpl.

\section{Preliminary Evaluation}
\label{sec:eval}

\subsection{Setup}
We ran an experiment in an OpenStack cluster that has 11 nodes (servers).
For \mysys, we carve out two storage nodes, i.e., nodes where \emph{cinder-volumes} are running, 
and other nodes are operated normally under a single OpenStack installation.
For the two nodes, we install the \mysys driver for \emph{cinder-volume} services
and expose raw local disk drives to \mysys system so that it can create a custom \stimpl 
based on the disks. The two nodes have 10 and 7 local HDDs respectively, each of which
has 1TB of storage capacity.

\subsection{Dynamic provisioning and meeting multiple requirements}
\begin{figure*}[t]
  \centering
  \includegraphics[viewport=55 230 750 510, scale=0.65, clip=true]{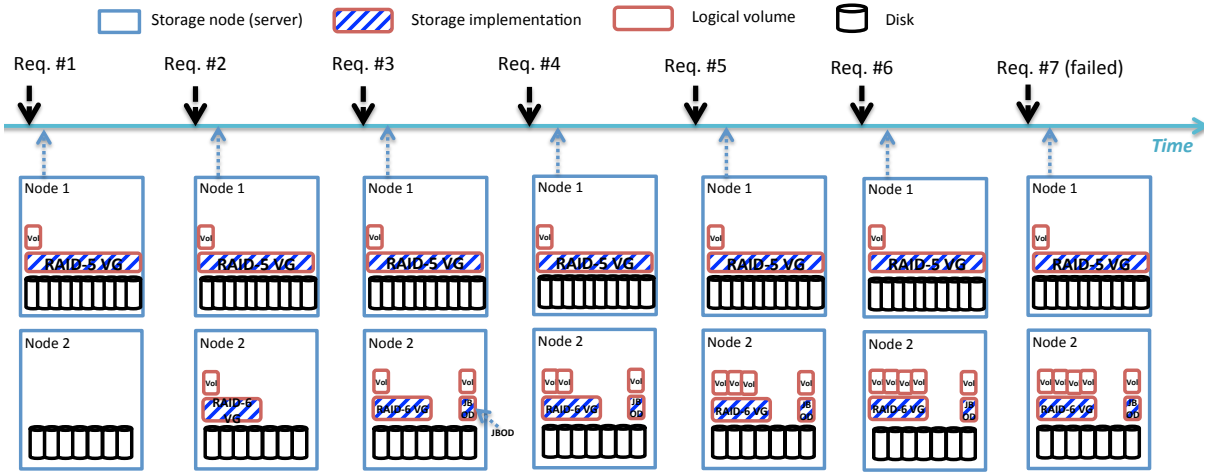}
  \vspace{-2ex}
  \caption {\mysys running on a real cluster: dynamic provisioning of a \stimpl 
		and admission control.}
  \label{fig:exp1}
\end{figure*}

\begin{table}[t]
\centering
\small
\begin{tabular}{|m{1.8cm}|m{6.2cm}|} \hline
\textbf{Volume Type} & \textbf{Property of \stimpl} \\ \hline
\hline
Type 1 & RAID-5 based on 10 disks.	\\ \hline
Type 2 & RAID-6 based on 4 disks. 100 IOPS. \\ \hline
Type 3 & JBOD based on a single disk. \\ \hline
\end{tabular}
\vspace{0.5ex}
\caption{
\label{tbl:voltype}Cinder volume types for the experiment}
\end{table}

\begin{table}[t]
\centering
\small
\begin{tabular}{|m{0.45cm}|m{0.75cm}|m{0.75cm}|m{0.75cm}|m{0.75cm}|m{0.75cm}|m{0.75cm}|m{0.75cm}|} \hline
\textbf{Req.}& \#1 & \#2 & \#3 & \#4 & \#5& \#6& \#7 \\ \hline
\hline
Spec.& Type~1 100G& Type~2 100G& Type~3 100G& Type~2 100G& Type~2 100G& Type~2 100G& Type~2 100G 	\\ \hline
\end{tabular}
\vspace{0.5ex}
\caption{
\label{tbl:request}Request Sequence (Type \# can be found in Table~\ref{tbl:voltype})}
\end{table}
We design an experiment that can show the benefits of dynamic provisioning 
of a \stimpl. Through the experiment, we aim to demonstrate the followings:

\begin{itemize}
	\item[*] Simultaneously meet a diverse set of tenant requests within a tight resource budget. 
			Each request has a different performance and reliability level.
	\item[*] Meet performance requirements of allocating an IOPS in a per-volume basis.
	\item[*] Perform a garbage collection for higher resource utilization. 
\end{itemize}

For the experiment, we configure three different \emph{volume type}s (Table~\ref{tbl:voltype})
and use the input sequence described in Table~\ref{tbl:request}. 
Each request is issued to the cluster every 90 seconds. 
The result is reported in Fig.~\ref{fig:exp1}.
The figure clearly shows that \mysys can dynamically create {\stimpl}s to satisfy 
heterogeneous requests (Req.\#1, \#2, and \#3). In addition,
it demonstrates that \mysys can allocate IOPS correctly (Req.~\#7 failed due to the budget constraint). 
In detail, our offline profiles tell us that we could use 200 IOPS (for 4k-block random read/write) per HDD
as a total performance budget. 
So, a RAID-6 based on 4 disks (on node 2) has 400 IOPS of a total performance budget, and
consequently the fifth trial with a type-2 request (Req.~\#7) failed; notice that 
a type-2 volume creation request comes with 100 IOPS.

After all 7 requests were processed, we also evaluated the garbage collection 
capability of \mysys system. In a software-RAID based implementation, we built 
a mechanism that could reclaim unused storage space, especially when there are 
no logical volumes in a given \stimpl for some amount of time. 
After deleting volumes, \mysys successfully reclaimed disks and each \stimpl 
returned to a raw disk pool.
Although not evaluated in this paper, it is worth noting that the garbage collection 
mechanism can be useful to other types of \stimpl, e.g., distributed storage systems such as Ceph. In case of Ceph, the garbage collection may trigger rebalancing operations to alleviate a skewed data distribution
in the cluster.

\subsection{Scheduling overhead}
\mysys has an additional intelligence for determining which host 
can satisfy a given requirement.
Nonetheless, it turns out the overhead incurred by \mysys system is negligible,
i.e., only about 5$\sim$6 milliseconds of latency will be added for \emph{cinder-scheduler}
to determine the eligibility of a storage host.

\section{Related Work}
\label{sec:related}

As cloud services boom, attention is drawn to providing users with block storage.
Amazon Elastic Block Store (EBS)~\cite{AmazonEBS} and OpenStack Cinder~\cite{OpenStackCinder} 
provide persistent block level storage volumes with virtual instances, usually via iSCSI~\cite{Satran2004rfc3720}.
For \emph{cinder-scheduler}, Yao et al. design and implement a new scheduling filter 
with the ability of IO throughput filtering and weighting to meet IOPS requirements~\cite{Yao2014SLA}, and propose a new block storage resource scheduling algorithm called MVBFD~\cite{Yao2015Multi}.


To the best of our knowledge, \mysys system is novel for its dynamic provisioning 
of a \stimpl in the area of cloud block storage.
Nonetheless, we can easily observe dynamic operations in modern distributed storage systems.
For instance, Ceph storage cluster can dynamically grow, shrink, and rebalance data 
along with the changes in underlying resources, 
e.g., \# of OSDs~\cite{Weil2006Ceph, Weil2007RADOS, Weil2006CRUSH}.
ASCAR~\cite{li15ASCAR} autonomously controls IO traffic from storage clients 
using a rule based algorithm called SHARP, which uses the congestion window 
and rate limit, in order to increase the bandwidth utilization and reduce speed variance.


\mysys implements a couple of simple, yet powerful, mechanisms to provide QoS. 
There exists an extensive body of research in the area of providing performance isolation 
and delivering guaranteed performance to tenants in today's virtualized datacenters.
%
Most of them are complementary to \mysys system in nature.
Gulati et al. proposes several approaches \cite{mClock, Gulati2009PARDA, Gulati2012Demand} 
and implements them in VMware ESX hypervisor.
mClock~\cite{mClock} proposes an IO scheduling algorithm in a hypervisor 
for per-VM QoS that supports reservation, limits, and shares.
PARDA~\cite{Gulati2009PARDA} provides proportional-share fairness 
to multiple hosts sharing a single storage array;
it detects overload via latency measurements, and then adjusts 
per-host issue queue lengths using flow control very similar to FAST TCP~\cite{FastTcp2006}.
SRP~\cite{Gulati2012Demand} proposes a hierarchical IO resource allocation system 
that support the logical grouping of related VMs into hierarchical pools; 
SRP controls IO reservations, limits and proportional shares at VM or 
pool level in an environment where VMs running multiple hosts are accessing the same storage.
Elnably et al.~\cite{Elnably2012EfficientQoS} proposes a scheduling algorithm 
called reward scheduling to provide QoS in a multi-tiered storage system 
that uses both SSD and HDD.
For key-value cloud storage, Pisces~\cite{Shue2012PerfIsolation} provides per-tenant 
weight fair sharing and performance isolation, and 
Pileus~\cite{Douglas2013ConsistencySLA} allows applications to use a choice of 
consistency guarantees with latency targets for consistency-based SLAs.

Tangentially related is a body of work that provides a QoS-aware 
block device interface to tenants in the cloud.  
%
Lin et al.~\cite{Lin2012Towards} proposes a block storage system design 
called FAST to minimize interference among tenants; 
read operations are redirected when different types of 
workloads (i.e., random and sequential reads) are co-located.
Mesnier et al.~\cite{Mesnier2011DiffStorage} proposes an IO classification 
architecture with a slightly-modified block interface 
in order to provide differentiated storage services.
Blizzard~\cite{Mickens2014Blizzard} extends the simple block interface 
for high IO performance, mapping each virtual driver to multiple 
backing physical disks, and providing high-level software abstractions 
such as replication and failure recovery.


%
\section{Conclusion and future work}
\label{sec:concl}

We present a system, called \mysys, that can dynamically provision 
an underlying storage implementation for cloud block storage services.
\mysys defers the storage provisioning process to the time at which a tenant
actually requests a storage space. The late binding can offer high flexibility to the cloud
service providers, thereby resulting in higher utilization.
\mysys implements two mechanisms, i.e., admission control and dynamic throttling, 
to provide QoS on performance among tenant applications. 
Through preliminary evaluation, we demonstrate that \mysys can satisfy
heterogeneous tenant requests even with considerably limited storage resources
and perform a proper admission control for allocating IOPS per volume.

Much work remains. We plan to integrate \mysys system with an emerging JBOF 
(Just-a-Bunch-Of-Flash) device and other distributed storage systems 
in the near future. 
Its QoS mechanism must be evaluated further thoroughly, even if our preliminary 
evaluation result looks promising.



{
\tiny
\bibliographystyle{abbrv}
\bibliography{mybib}
}

\end{document}